# The colloidal nature of complex fluids enhances bacterial motility

Shashank Kamdar[1], Seunghwan Shin[1], Premkumar Leishangthem[2], Lorraine F. Francis[1,*], Xinliang Xu[2,3,†], and Xiang Cheng[1,‡]

[1] Department of Chemical Engineering and Materials Science, University of Minnesota, Minneapolis, MN 55455, USA

[2] Complex Systems Division, Beijing Computational Science Research Center, Beijing 100193, China

[3] Department of Physics, Beijing Normal University, Beijing 100875, China

* lfrancis@umn.edu

† xinliang@csrc.ac.cn

‡ xcheng@umn.edu

**The natural habitats of microorganisms in the human microbiome, ocean and soil ecosystems are full of colloids and macromolecules. Such environments exhibit non-Newtonian flow properties, drastically affecting the locomotion of microorganisms[1-5]. Although the low-Reynolds-number hydrodynamics of swimming flagellated bacteria in simple Newtonian fluids has been well developed[6-9], our understanding of bacterial motility in complex non-Newtonian fluids is less mature[10,11]. Even after six decades of research, fundamental questions about the nature and origin of bacterial motility enhancement in polymer solutions are still under debate[12-23]. Here, we show that flagellated bacteria in dilute colloidal suspensions display quantitatively similar motile behaviors to those in dilute polymer solutions, in particular, a universal particle-size-dependent motility enhancement up to 80% accompanied by a strong suppression of bacterial wobbling[18,24]. By virtue of the hard-sphere nature of colloids, whose size and volume fraction we vary across experiments, our results shed new light on the long-standing controversy over bacterial motility enhancement in complex fluids and suggest that polymer dynamics may not be essential for capturing the phenomenon[12-23]. A physical model that incorporates the colloidal nature of complex fluids quantitatively explains bacterial wobbling dynamics and mobility enhancement in both colloidal and polymeric fluids. Our findings contribute to the understanding of motile behaviors of bacteria in complex fluids, which are relevant for a wide range of microbiological processes[25] and for engineering bacterial swimming in complex environments[26,27].**

We use fluorescently tagged wild-type *Escherichia coli* in our experiments (Methods), which execute the classic "run-and-tumble" motion and show wobbling in Newtonian fluids (Fig. 1a inset)[8,24]. As a model organism, *E. coli* is the most widely used flagellated bacteria in motility studies[12-14,16,18-20]. We suspend *E.*

*coli* cells in a minimal phosphate motility buffer that does not allow further growth of bacteria during experiments. The bacterial sample is then mixed with a colloidal suspension of controlled particle radius $R$ to achieve a desired colloidal volume fraction $\phi$. We use commercial polystyrene (PS) and polymethyl methacrylate (PMMA) colloids, which are extensively washed and dialyzed before use. The bacterial concentration is fixed at $5.2 \times 10^8$ cells/ml, far below the onset of the collective swimming of bacterial suspensions[28,29]. We inject the bacteria-colloid mixture into a closed 150-μm-deep polydimethylsiloxane (PDMS) microfluidic channel. PDMS is oxygen permeable, so that the channel is well oxygenated during a single motility measurement over 15 minutes (Extended Data Fig. 1a). Using confocal microscopy, we image bacteria 30 μm above the channel bottom to avoid the interaction of bacteria with the surface[30]. For each experiment at given $R$ and $\phi$, we track more than 1000 independent bacterial trajectories to obtain the average swimming speed and wobble angle of bacteria (Figs. 1a and b). To calibrate our experimental protocol, we measure the swimming speed of bacteria in Ficoll 400 polymer solutions, which shows quantitatively the same results as those reported previously (Extended Data Fig. 1b)[16].

Figure 1c shows the normalized average bacterial swimming speed $V/V_0$ as a function of $\phi$ for colloids of different $R$, where $V_0$ is the average swimming speed of bacteria in the buffer without colloids. Bacterial swimming speeds are non-monotonic with increasing $\phi$ and exhibit motility enhancement in dilute suspensions. $V/V_0$ reaches a maximum, $V_{max}/V_0$, at low $\phi < 0.04$ and then decreases at higher $\phi$. Similar non-monotonic trends have also been observed for bacterial swimming in polymer solutions with increasing polymer concentrations, where motility enhancement was found in dilute solutions[12,13,16,18,21]. We also plot the probability distributions of the normalized bacterial swimming speeds in suspensions of 500-nm colloids at different $\phi$ (Fig. 1d). The invariance of the distributions shows that the increase of the average speed is due to a uniform increase of speeds of all bacteria, instead of a small fraction of bacteria.

More importantly, we find that the degree of the motility enhancement depends on the size of colloids. $V_{max}/V_0$ increases monotonically with $R$ within the range of our experiments (Fig. 2). A motility enhancement as large as 80% is observed for large 500-nm colloids. Remarkably, when we compare the size-dependent bacterial motility enhancement in colloidal suspensions with that in polymer solutions, all the data in the literature with well-defined molecular weights collapse into a universal curve. Here, we estimate the size of a polymer coil using its hydrodynamic radius (Methods)[31]. Most previous studies of bacterial swimming in polymer solutions focused on the dependence of bacterial speeds on the concentration or the viscosity of polymer solutions[12-14,16,18-20]. Hence, experiments using polymers of small molecular weights did not reveal obvious speed enhancement at the same concentration or viscosity where other studies using large polymer molecules showed clear speed enhancement[16,17,20]. This controversy over the existence of bacterial motility enhancement can be resolved by explicitly considering the size of polymer molecules. Although the size dependence can be inferred from recent experiments on polymers of different molecular weights[16,18] and has been discussed in a recent theory[17], our experiments provide the clear evidence on the particle-size dependence of motility enhancement. The well-controlled size of colloids allows us to explore a wide range of particle sizes using colloids of similar properties, which is hard to achieve with polymeric systems.

The similarity of the mobile behaviors of bacteria in colloidal suspensions and in polymer solutions goes beyond the size-dependent motility enhancement. In fact, all the reported features of bacterial swimming in dilute colloidal suspensions are quantitatively similar to those in dilute polymer solutions. Bacterial tumbling rates decrease with increasing colloidal volume fractions or polymer concentrations, leading to smoother and less tortuous trajectories at high concentrations in both fluids (Figs. 1a and b)[18]. Analyses of the rotational diffusivity, the mean run time and tumble time of bacteria confirm their quantitative similarities (Extended Data Fig. 2). Note that tumbling events accounts only about 10% of the total time of bacterial swimming[8]. Thus, although the decrease of tumbling rates contributes to the motility enhancement, it is insufficient to explain the 80% increase in bacterial swimming speed. Moreover, bacterial swimming speeds are nearly linearly anticorrelated with the wobble angles of bacteria: fast bacteria wobble less. This anticorrelation is observed not only between the average swimming speed and the average wobble angle of bacteria in suspensions of different $\phi$ (Fig. 3), but also between the swimming speed and the wobble angle of individual bacteria at each given $\phi$ (Extended Data Fig. 3). The trend is again quantitatively similar to that observed for bacterial swimming in polymer solutions[18,32].

Together, seven bacterial swimming features share quantitative similarities in dilute polymer solutions and colloidal suspensions, which are summarized in Sec. 1 of Supplementary Information (SI). This comprehensive list strongly suggests a common origin of the enhanced motility of flagellated bacteria in both types of complex fluids. The observation challenges the existing theories on bacterial motility enhancement, which have been developed based on various aspects of polymer dynamics, irrelevant to our colloidal system. (*i*) Our colloids are non-attractive, which cannot form gel network at the low $\phi$ of our experiments. Thus, the widely-circulated mechanism based on the formation of gel network with bacteria-sized pores cannot explain the motility enhancement in our experiments[14,15]. (*ii*) The viscosity of colloidal suspensions with $\phi < 0.04$ follows the Einstein equation and is less than 10% above that of the buffer[33]. The degree of shear thinning in such low-$\phi$ suspensions is too weak to induce the 80% motility enhancement within different two-fluid models[16,20,21,23]. (*iii*) Our colloids are hard with Young's moduli of 2-4 GPa, which cannot be deformed by the shear flow of bacterial flagella. Thus, the recent proposal that the motility enhancement arises from the normal stress difference induced by the stretching of polymer coils near bacterial flagella is not applicable either[18,22]. (*iv*) We have extensively dialyzed and washed colloids before use. There are no metabolizable organic impurities in our suspensions that can lead to motility enhancement[16].

Our experiments with dilute hard-sphere colloidal suspensions demonstrate the importance of the colloidal nature of complex fluids in controlling bacterial swimming. Inspired by the finding, we develop a simple model based on the hydrodynamic interaction between individual bacteria and colloids, which quantitatively explains the enhanced motility of flagellated bacteria in both colloidal and polymeric fluids. Specifically, we tackle the decades-old problem in two steps. First, we show the hydrodynamic interaction of a bacterium with a colloid or a polymer coil reduces bacterial wobbling and results in the size-dependent motility enhancement shown in Fig. 2. Second, we propose a mechanism of bacterial wobbling, which quantitatively predicts the anticorrelation between bacterial swimming speeds and wobble angles shown in Fig. 3. We summarize the main results of our model below with the detailed derivation given as SI.

When a sphere translates along a solid surface in a low-Reynolds-number flow, the hydrodynamic interaction between the sphere and the surface induces the rotation of the sphere as if it experiences an effective torque from fluid, an effect often referred to as the boundary-induced torque (BIT)[34,35]. We consider the swimming of a bacterium near a colloid of radius $R$ with a velocity $V_b$. $V_b = CV_0$ is the velocity of the bacterium along its helical trajectory with $C > 1$ a proportional constant. The projection of $V_b$ along the helical axis of the trajectory gives the swimming speed $V$ measured in experiments (Extended Data Fig. 4a). The BIT on the bacterial body due to the presence of colloids can be calculated under the lubrication approximation,

$$\tau_{BIT} = \frac{\pi}{5}\mu r_b^2 V_b \frac{\gamma(17+8\gamma)}{(1+\gamma)^2} \ln\frac{r_b(1+\gamma)}{2d}, \tag{1}$$

where $\gamma \equiv R/r_b$ with $r_b = 1$ μm the characteristic radius of bacterial body, $\mu = 10^{-3}$ Pa·s is the buffer viscosity, and $d$ is the minimum surface distance between the colloid and the bacterial body. We choose $d \approx 50$ nm based on previous experiments[36]. The lubrication approximation applies when $(r_b, R) > d$. The torque-free condition requires that the BIT is balanced by the restoring torque of the bacterial body induced by the change of its wobble angle $\theta$ away from the equilibrium wobble angle $\theta_0$ without colloids, $\tau_r = -K(\theta - \theta_0)$. Here, we assume a linear elastic response of the flagellar hook with a bending stiffness $K$. Using the empirical linear relation between bacterial speeds and wobble angles, $\theta = a_1 - b_1 V$, as an input (Fig. 3b), the torque balance between $\tau_{BIT}$ and $\tau_r$ gives

$$\frac{V}{V_0} = 1 + \frac{\pi}{5}\frac{C}{K}\frac{\mu r_b^2}{b_1}\frac{\gamma(17+8\gamma)}{(1+\gamma)^2}\ln\frac{r_b(1+\gamma)}{2d}, \tag{2}$$

where $b_1 = 3.61 \times 10^{-2}$ s/μm is the slope of the linear fitting. $C = 2.67$ is fixed by the condition $V_b = V(\theta = 0)$. Equation (2) quantitatively agrees with experiments with $K = 7.16 \times 10^{-19}$ N·m (the black line in Fig. 2), matching well with the estimate of the bending stiffness of the flagellar hook at $5 \times 10^{-19}$ N·m[37]. This parameter-free description of bacterial motility enhancement is generic, as the BIT arises from the no-slip boundary condition on the surface of spherical particles, which applies to both colloids and polymer coils. It has been well established in polymer science that the random coil of a polymer molecule in dilute solutions behaves hydrodynamically like a non-draining sphere with the size equal to its hydrodynamic radius[31].

While the simple generic calculation successfully predicts the motility enhancement without recourse to the detailed kinematics of bacterial locomotion, it cannot address the important question: why does the decrease of wobble angles lead to the increase of bacterial speeds? To answer the question, we further examine the origin of bacterial wobbling. The trajectory of bacteria is generally helical in 3D, which manifests as the wobbling of bacterial body in the 2D projection of optical microscopy (Extended Data Fig. 4a)[24,32,38]. Different from the origin of the helical trajectory of microalga with actively beating flagella[39-41], the helical trajectory of flagellated bacteria arises mainly because of the asymmetric off-axial geometry of the passively rotating flagellar bundle[24,32]. Nevertheless, a minimal model based on the simple asymmetric geometry could not reproduce helical trajectories with pitches larger than 4 μm, which are frequently observed for bacterial swimming with a single flagellar bundle[24]. Moreover, a model considering the change of the drag coefficients due to the off-axial flagellar bundle predicted only less than 8% motility enhancement when the wobble angle decreases over the full range from 90° to 0°.[20] Thus, the current understanding of bacterial wobbling is still incomplete. Here, we propose a new model of bacterial wobbling. In our model, bacterial body and flagellar bundle rotate at angular velocities, $\omega_b$ and $\omega_t$, respectively (Extended Data Fig. 4b). While $\omega_b$ and $\omega_t$ are correlated by the balance of the flagellar motor torque[42], they are generally not collinear, which results in a rigid-body rotation of the bacterium as a whole, $\omega_{cm}$, and, therefore, bacterial wobbling. Using the torque-free and force-free conditions, we calculate $\omega_{cm}$ and bacterial velocity $V_b$ tangential to its helical trajectory. The wobble angle $\theta$ is formed by $\omega_b$ and $-\omega_{cm}$ (supplementary equation (32)), whereas the measured swimming speed of the bacterium along the helical axis of wobbling trajectories is $V = |V_b \cdot \omega_{cm}|/|\omega_{cm}|$ (supplementary equation (34)). Without fitting parameters, our model quantitatively predicts not only the anticorrelation between bacterial speeds and wobble angles (Fig. 3b), but also the probability distribution of the pitch of bacterial helical trajectories (Extended Data Fig. 5). Particularly, the model shows that, with a modest decrease of the wobble angle of ~ 20°, a strong speed enhancement of 80% can be achieved (Extended Data Fig. 4d). The model also supports a symmetry breaking mechanism, where the hydrodynamic interaction between a bacterium and a colloid preferentially reduces the wobbling of the bacterium and therefore enhances its motility (Extended Data Figs. 6 and 7).

Thus, our study combining experiments and theory solves two long-standing problems of bacterial locomotion, i.e., the origin of bacterial motility enhancement in complex fluids and the mechanism of bacterial wobbling in simple Newtonian fluids. We show that the hydrodynamic interaction between individual bacteria and the colloidal component of complex fluids mediated by the background Newtonian fluid reduces bacterial wobbling, which in turn enhances bacterial motility as the consequence of their wobbling dynamics. The rheology of polymer solutions seen as continuum media plays only a minor role in this process. Indeed, we can visualize the effect of the discrete colloidal interaction on bacterial swimming by imaging bacterial trajectories near a colloid held in an optical trap. The motility enhancement and the suppression of bacterial wobbling can be identified close to the colloid (Fig. 4 and Supplementary Video). Unique polymer dynamics become important in semi-dilute and concentrated solutions[31], where the viscoelasticity, normal stress difference and shear thinning of polymer solutions likely affect the decrease of bacterial motility at high polymer concentrations[16].

It is an open question how the maximum volume fraction, $\phi_{max}$, where the mobility enhancement peaks, depends on the size of colloids or polymer coils. Experimentally, it is hard to determine $\phi_{max}$ accurately due to the difficulty in measuring absolute volume fractions[43]. $\phi_{max}(R)$ scatters at low concentrations without a clear trend (Extended Data Fig. 8). Theoretically, $\phi_{max}$ should depend on the relaxation of a bended flagellar hook between successive encounters of a bacterium with multiple colloids and also the steric bacterium-colloid interaction that becomes increasingly important at high $\phi$. The question is beyond the scope of our simple model and poses a challenge for future studies.

**Figures:**

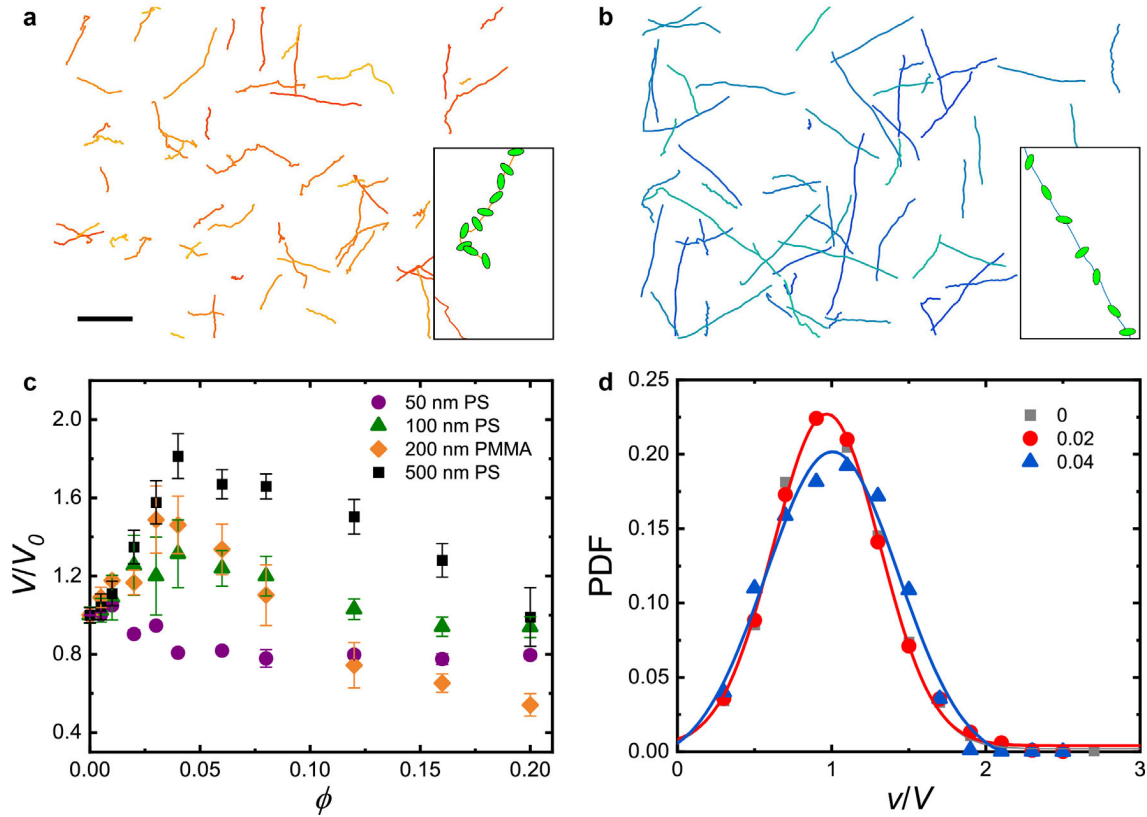

**Fig. 1 | Bacterial swimming in colloidal suspensions. a,b**, Representative bacterial trajectories in the motility buffer (**a**) and in a suspension of polystyrene colloids of radius $R$ = 500 nm at volume fraction $\phi$ = 0.04 (**b**). Scale bar: 20 μm. Insets: Zoomed-in view of bacterial trajectories with instantaneous bacterial orientation indicated. The length of bacteria is 3 μm. Bacteria show longer and straighter trajectories and less wobbling in the colloidal suspension, qualitatively similar to those observed in dilute polymer solutions (see Fig. 1 of Ref. [18]). **c**, The average swimming speed of bacteria $V$ as a function of the volume fraction $\phi$ of suspensions of colloids of different $R$. $V$ is normalized by the average swimming speed of bacteria in the buffer without colloids $V_0$ = 13 ± 2 μm/s. $R$ is indicated in the figure. **d**, Probability distribution functions (PDFs) of the normalized bacterial speed, $v/V$, for suspensions of colloids of $R$ = 500 nm at three different $\phi$. $v$ is the swimming speed of individual bacteria, whereas $V$ is the average swimming speed of all bacteria at the corresponding $\phi$. $V = V_0$, 17.5 μm/s and 23.5 μm/s for $\phi$ = 0, 0.02 and 0.04, respectively. The distributions are fitted with Gaussian functions (solid lines), reflecting the intrinsic cell-to-cell variation in swimming speeds. The fitting functions at $\phi$ = 0 and 0.02 overlap well.

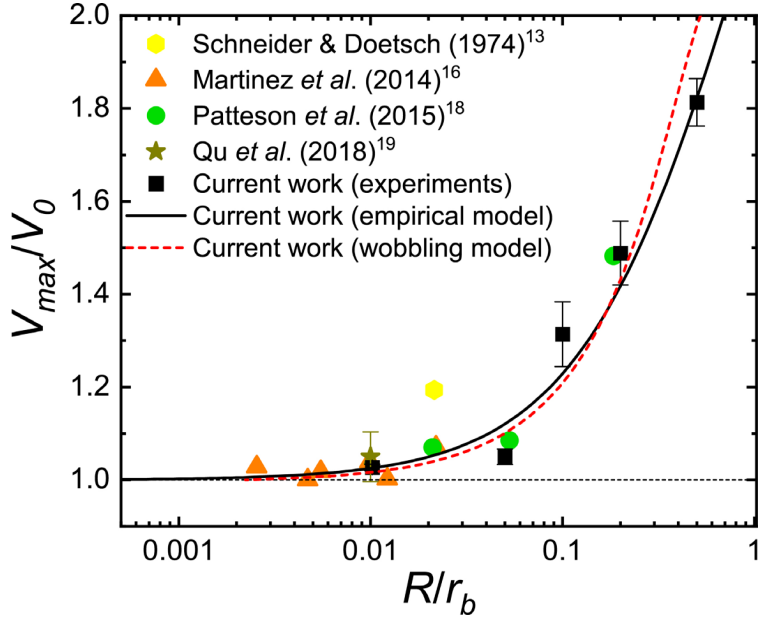

**Fig. 2 | Size-dependent bacterial motility enhancement.** The normalized maximum bacterial swimming speed, $V_{max}/V_0$, as a function of the radius of colloids or the hydrodynamic radius of polymer coils, $R$. $R$ is normalized by the characteristic radius of bacterial body $r_b = 1$ μm. Data from previous experiments on polymer solutions are referenced in the figure. The black solid line is the prediction of equation (2) based on the empirical linear fitting (the dashed line in Fig. 3b), whereas the red dashed line is the prediction of our wobbling model (SI Sec. 3). $V_{max}(R)$ reaches the maximum when the wobble angle $\theta$ approaches zero. The empirical model predicts the maximum $(V_{max}/V_0)_{max} = V_b/V_0 = 2.67$ occurring at $R/r_b = 2.22$. The wobbling model predicts $(V_{max}/V_0)_{max} = 2.09$ occurring at $R/r_b = 3.68$.

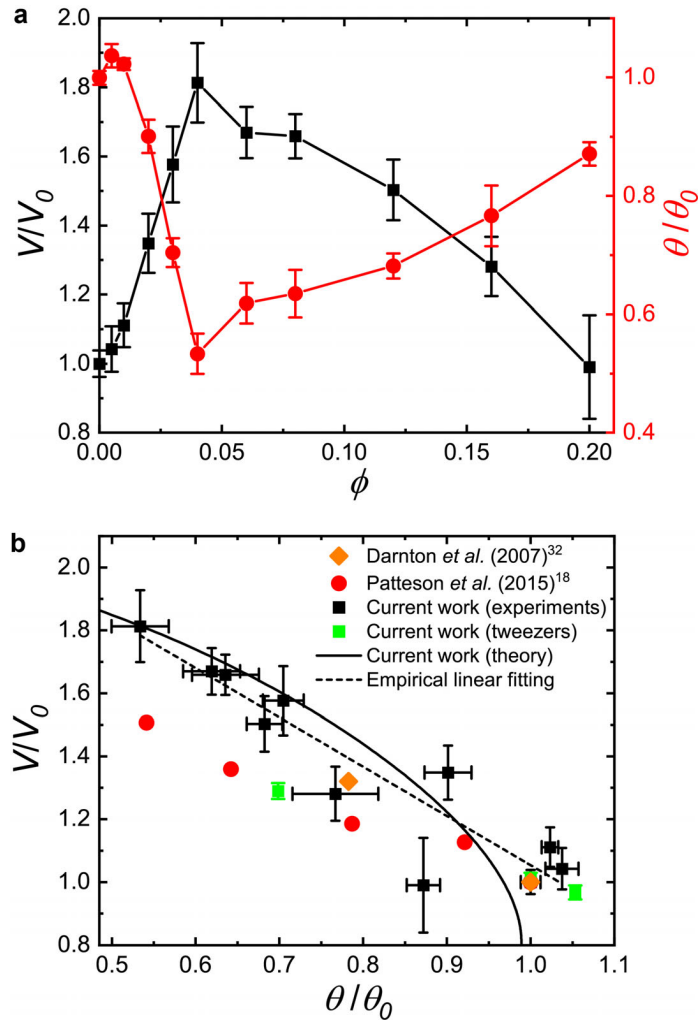

**Fig. 3 | Anticorrelation between the swimming speed and wobble angle of bacteria. a**, The swimming speed $V$ and the wobble angle $\theta$ of bacteria as a function of the volume fraction of colloidal suspensions $\phi$. The radius of colloids is $R = 500$ nm. $V$ is normalized by the average swimming speed of bacteria in the buffer $V_0$. $\theta$ is normalized by the average wobble angle of bacteria in the buffer $\theta_0 = 45.1° \pm 0.5°$. **b**, $V/V_0$ versus $\theta/\theta_0$. Data from previous experiments on polymer solutions are referenced in the figure. Our measurements with optical tweezers show the same trend (green squares). The dashed line is an empirical linear fitting used in equation (2), whereas the solid line is the prediction of our model of bacterial wobbling (see also Extended Data Fig. 4d). Quantitatively similar anticorrelation has also been observed between the swimming speed and wobble angle of individual bacteria in a given suspension of fixed $\phi$ (see Extended Data Fig. 3).

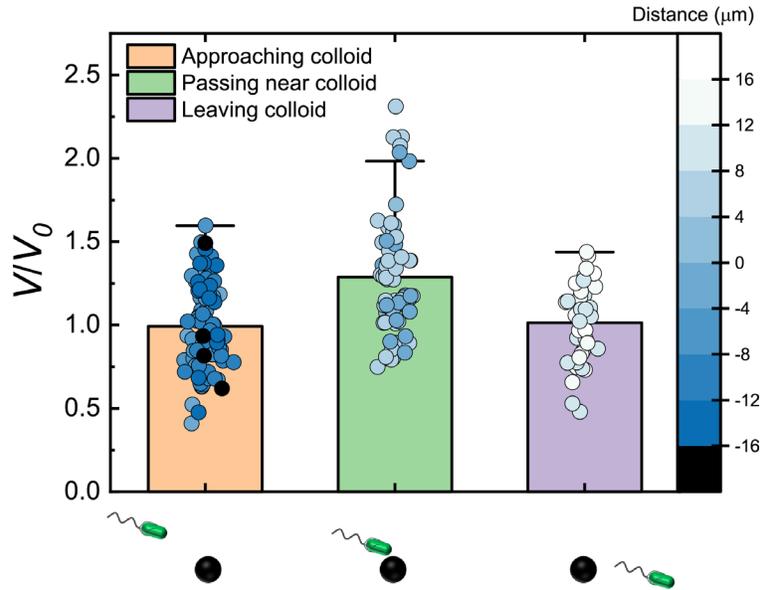

**Fig. 4 | Enhanced motility of bacteria near a colloid.** The colloid of radius $R = 1.5$ μm is held in an optical trap (Methods). The swimming speeds of bacteria $V$ approaching, passing near, and leaving the colloid are measured. The approaching regime is defined when the distance between the center of bacterial body and the center of the colloid $d_0 < -3$ μm, whereas the leaving regime is defined when $d_0 > 8$ μm. $V$ is normalized by the average swimming speed of bacteria far away from the colloid $V_0$. Symbols represent the results of individual measurements, whereas the histogram shows the average speeds in the three regimes. $V/V_0 = 1.29$ in the passing-near regime and $V/V_0 = 1$ in the approaching and leaving regimes. The symbols are color-coded according to $d_0$. The average wobble angles versus the average swimming speeds in the three regimes are shown in Fig. 3b. A representative trajectory of a bacterium near the colloid is shown in Supplementary Video.

## Methods

### Bacteria

We used a fluorescently tagged wild-type *E. coli* strain (BW 25113) engineered to carry PKK_PdnaA-GFP plasmids. The bacteria from stock solution were grown overnight (12-16 hours) at 35 °C in nutrient rich terrific broth (TB: tryptone 11.8 g/L, yeast extract 23.6 g/L, glycerol 4 ml/L) supplemented with 0.1% (v/v) ampicillin (antibiotic, 100 mg/L). A small amount of this overnight culture was diluted (1:100) in TB to grow a fresh culture for 6.5 hours at 30 °C in the exponential phase. Motile cells were harvested by centrifugation at 800*g* for 5 min. After centrifugation, the growth medium was discarded, and cells were washed twice with motility buffer (MB: 0.01M potassium phosphate, 0.067M NaCl, $10^{-4}$ M EDTA, pH 7.0). Motile cells were carefully transferred to a tube and the concentration of the cells in MB was measured using a biophotometer via $OD_{600}$. Subsequent motility measurements were performed at 22 °C. Bacteria remain active with constant motility for at least 3 hours in the tube. After bacterial suspensions were injected into the closed PDMS microchannels, we conducted our experiments within 15 min, within which the average bacterial swimming speed stays constant (Extended Data Fig. 1a).

### Colloids

Polystyrene (PS) and polymethyl methacrylate (PMMA) colloidal particles of different sizes were bought from Bangs Laboratories. The polydispersity index (PDI) of particles is less than 0.05. To remove the trace amount of impurities, we cleaned the particles seven times in deionized (DI) water via alternating centrifugation and resuspension. PS particles of radius 100 nm were further cleaned using a Float-A-Lyzer dialysis device (MWCO: 300kDa, Repligen), which removed small impurities of size 20 nm and below. The average swimming speeds of bacteria were quantitatively the same before and after dialysis. We measured the volume fraction of colloids using NanoSight LM-10 (Malvern Panalytical), as well as by drying and weighing the colloids[43]. Absence of colloidal aggregation was verified using bright-field microscopy. The colloidal suspension was mixed with the bacterial suspension to achieve the desired colloidal volume fraction $\phi$. Bacterial concentration in the final mixture was low and fixed in our experiments at $5.2 \times 10^8$ cells/ml to ensure negligible cell-cell interactions.

### PDMS microchannels

Polydimethylsiloxane (PDMS) microchannels were fabricated following the standard soft lithography process. The master mold was prepared by spin coating photoresist SU-8 2050 (Micro Chem) on a 4-inch silicon wafer at 1500 rpm for 30 s. The wafer was soft baked at 65 °C for 5 min, and then photoresist was exposed through a photomask. The process was followed by a post-exposure bake at 95 °C for 30 min. The mold was developed in an ultrasonic bath and the height of features was controlled by etching duration.

Subsequently, the channel depth was measured using a profilometer (KLA-Tencor P7). A negative replica of the SU8 master mold was cast in PDMS. A coverslip was base-washed (NaOH 1M), which made the surface hydrophilic and reduced the adhesion of bacteria on the surface. The PDMS surface and the glass coverslip were plasma treated and adhered. The channel was left at 75 °C for 15 min to enhance the bonding between the glass coverslip and the PDMS surface. The final microchannel had a dimension of 2 cm (L) by 1.8 mm (W) by 150 μm (H). Bacterium-colloid mixtures were slowly deposited into the microchannel to avoid flagellar damage caused by the shear flow. The mean square displacements of colloids in water without bacteria showed normal Brownian diffusion in the microchannel without detectable drift.

**Confocal microscopy and optical tweezers**

We performed confocal microscopy using an inverted confocal microscope with a 488 nm laser (Nikon Ti-Eclipse). Images were captured by an Andor Zyla sCMOS camera 30 μm above the surface of the coverslip to avoid the influence of bacteria-surface interactions[30]. In order to account for any spatial heterogeneity, 1-min videos were recorded for each suspension at three different locations of the same height near the center of the microchannel. The videos were taken at 30 frames per second using a Nikon Plan Apo 60× oil objective (NA = 1.4) with a field of view of 200 μm by 100 μm. The procedure yielded more than 1000 independent bacterial trajectories for each $\phi$, ensuring statistical significance of our measurements.

We used optical tweezers (Aresis Tweez-250si) to set up an optical trap, which held a large 1.5-μm-radius colloid 15 μm above the coverslip surface. The tweezers used an infrared laser of 1064 nm with a laser power of 10 mW, which weakly held the colloid. The potential of the trap decayed substantially outside the radius of the large colloid and, therefore, did not affect the motions of bacteria passing near the colloid.

**Swimming speed and wobble angle**

We used a custom Matlab code for *E. coli* cell identification and body-wobble-angle measurements. A standard particle tracking algorithm was then used to construct the trajectories of all the bacteria in the videos[44]. To reduce noise, we removed short trajectories where bacteria appeared less than ten consecutive frames in the focal plane. Dead and immobile bacteria were also removed by imposing a speed cutoff at 4 μm/s. The speed of bacteria was calculated from the ratio of the displacement of the center of bacterial body between two consecutive frames to the inverse frame rate. Using the central finite difference for the speed calculation yielded quantitatively the same results. The average swimming speed of bacteria was finally obtained by taking the temporal and ensemble average of all the trajectories within the videos. The results thus obtained for bacterial swimming in the solutions of Ficoll 400 are quantitatively the same as the speed measurements reported in previous studies using differential dynamic microscopy and dark-field flicker microscopy (Extended Data Fig. 1b)[16].

To extract the wobble angle of bacteria, we first measured the temporal evolution of the orientation of bacterial body, $\delta(t)$, with respective to the mean direction of bacterial motion given by the unit tangential vector along the bacterial trajectory at time $t$ (Fig. 1a and b inset). The amplitude of the oscillation of $\delta(t)$ averaged over all bacterial trajectories was taken as the wobble angle of bacteria, $\theta$.

**Hydrodynamic radius and concentration of polymer solutions**

Previous experimental studies on the motility enhancement of flagellated bacteria in polymer solutions focused on the dependence of bacterial swimming speed on the concentration or the shear viscosity of polymer solutions[12-14,16,18,20]. Few studies have explicitly reported the hydrodynamic radius, $R_h$, of polymer molecules. Hence, we had to extract $R_h$ of polymer molecules based on the average molecular weight of polymer and the viscosity of dilute polymer solutions published in these studies. Specifically, the viscosity of dilute polymer solutions can be expressed via a virial expansion $\eta = \eta_s(1 + [\eta]c + O(c^2))$, where $\eta$ is the viscosity of polymer solutions, $\eta_s$ is the solvent viscosity, $c$ is the polymer concentration and $[\eta] = \lim_{c \to 0}\left(\frac{\eta - \eta_s}{\eta_s c}\right)$ is the intrinsic viscosity[31]. The intrinsic viscosity is related to the hydrodynamic volume of polymer molecules $V_h$ through $[\eta] = 2.5 N_{AV} V_h / M$, where $N_{AV}$ is the Avogadro constant and $M$ is the polymer molecular weight[31]. From $V_h = 4\pi R_h^3/3$, we have

$$R_h = \left(\frac{3[\eta]M}{10\pi N_{AV}}\right)^{1/3}. \tag{3}$$

$[\eta]$ can be directly extracted from the reported rheological data of polymer solutions in the dilute limit, where a linear fitting between $(\eta - \eta_s)/\eta_s$ and $c$ yields $[\eta]$ as the slope of the fitting. As an example of our analysis, data from the study of Patteson and co-workers are shown and analyzed in Extended Data Fig. 1c[18]. Given the average molecular weight, we can then determine the average $R_h$ from equation (3). Note that we can extract $R_h$ only from those studies, which reported both the rheological properties and the average molecular weights of polymer. The hydrodynamic radius of polymers such as Methocel™ cannot be estimated due to their composite nature with wide molecular weight distributions[12,14,20]. Figure 2 included all the data from previous studies, where we could successfully extract $R_h$.

Note that different types of polymers with the same molecular weight (or the same number of repeat units) can have completely different hydrodynamic radii and, therefore, have drastically different effects on the swimming speed of bacteria. In addition to the molecular weight, the size of a polymer coil also depends on the molecular structure and the solvent quality. For example, branched polymers such as Ficoll have relatively small sizes even at high molecular weights, which then have a weak effect on bacterial motility enhancement. On the other hand, linear chains such as methylcellulose (MC) and polyvinylpyrrolidone (PVP) have much larger sizes at the same molecular weight and thus have a much stronger effect on the motility enhancement.

Finally, we estimated the volume fraction of polymer solutions where the motility enhancement is strongest, $\phi_{max}$, based on the overlap concentration of the solutions. Specifically, $\phi_{max} = (c_{max}/c^*)\phi^*$, where $c_{max}$ is the polymer concentration at the peak motility, $c^*$ is the overlap concentration and $\phi^*$ is the polymer volume fraction at $c^*$. We obtained $c_{max}$ from the published results and approximated $\phi^* \approx 0.64$ as the upper limit. Note that we did not use $R_h$ for the volume fraction calculation. As $\phi \sim R_h^3$, even a small error in the calculation of $R_h$ leads to a large uncertainty in $\phi$. Measurements and comparison of the absolute $\phi$ of different experiments are challenging even for suspensions of nearly monodisperse hard-sphere colloids[43], let alone polymer solutions with wide molecular weight distributions. As a result, experiments from different laboratories on the same polymer solution could yield quite different $\phi_{max}$ (e.g., data on the solution of Ficoll 400 at $R/r_b = 10^{-2}$ in Extended Data Fig. 8).

**Acknowledgments** We thank Pranav Agrawal, Supriya Ghosh, Younjun Kim, Zhengyang Liu, Roshan Patel and Gregorius Pradipta for help with experiments and data analysis and Shuo Guo, Panayiotis Kolliopoulos, Eric Lauga, Yi Man and Zijie Qu for fruitful discussions. We would also like to acknowledge the late Prof. Howard Berg, who has answered our questions on *E. coli* mutant strains. This research is supported by the IPRIME program of University of Minnesota, and by the US National Science Foundation CBET-1702352 and 2028652. X.X. acknowledges the financial support from National Natural Science Foundation of China No. 11974038, and No. U1930402. Portions of this work were conducted in the Minnesota Nano Center, which is supported by the US National Science Foundation through the National Nano Coordinated Infrastructure Network (NNCI) under Award Number ECCS-2025124.

**Author contributions** S.K., L.F. and X.C. designed the research. S.K. performed the experiments. S.K., S.S., L.F. and X.C. discussed and analyzed the experimental data. X.X. proposed the model and conducted the analytical calculations with input from X.C.. P.L. and X.X. performed the simulation of Extended Data Fig. 7. X.C. conceived the project. L.F., X.X. and X.C. supervised the project. S.K. and X.C. co-wrote the manuscript. All authors discussed and commented on the manuscript.

**Competing interests** The authors declare no competing interests.

**Additional information**
**Correspondence and requests for materials** should be addressed to Xiang Cheng or Xinliang Xu.
**Supplementary Information** is available for this paper.

**Data availability**
All data needed to evaluate the conclusions herein are available from the University of Minnesota Data Repository: https://doi.org/10.13020/nfr5-te36.

**Code availability**
The codes used for tracking bacterial swimming speeds and wobble angles are available from the University of Minnesota Data Repository: https://doi.org/10.13020/nfr5-te36.

**Extended Data Figure legends:**

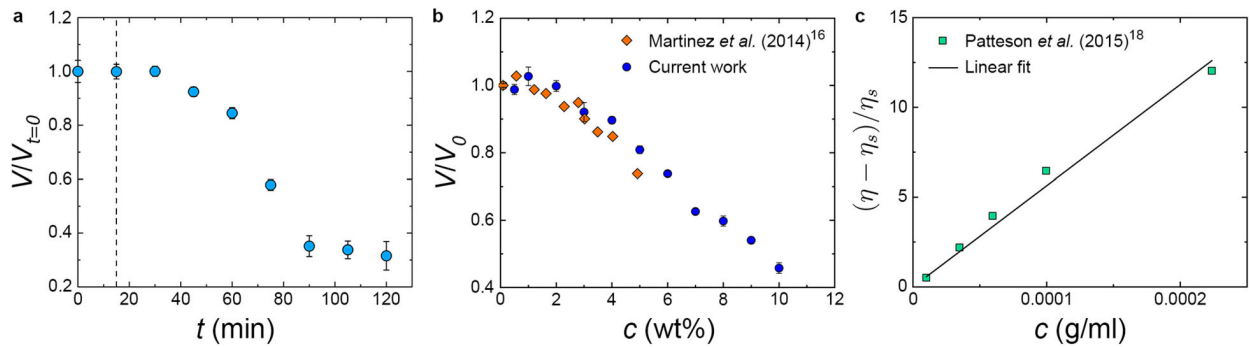

**Extended Data Fig. 1 | Experimental methods. a**, The average swimming speed of bacteria $V$ as a function of time after a bacterial suspension is injected into a PDMS microchannel. $V$ is normalized by the average speed at time $t = 0$. The vertical dashed line indicates the maximum measurement time of our experiments of 15 min. **b**, Comparison of the average swimming speed of bacteria $V$ in suspensions of Ficoll 400 of increasing concentrations $c$ from our experiments with that from a previous study. $V$ is normalized by the average swimming speed in the pure buffer $V_0$. $c$ is the weight percentage concentration. Blue circles are from our experiments, whereas orange diamonds are from Ref. [16]. **c**. Determination of the intrinsic viscosity of dilute solutions of carboxymethyl cellulose (CMC) of average molecular weight 700,000. The viscosities of the solutions $\eta$ as a function of polymer concentrations $c$ are extracted from Ref. [18]. $\eta_s$ is the solvent viscosity. $c$ is in unit of g/ml. The slope of the linear fit gives $[\eta]$ = 56344.1 ml/g.

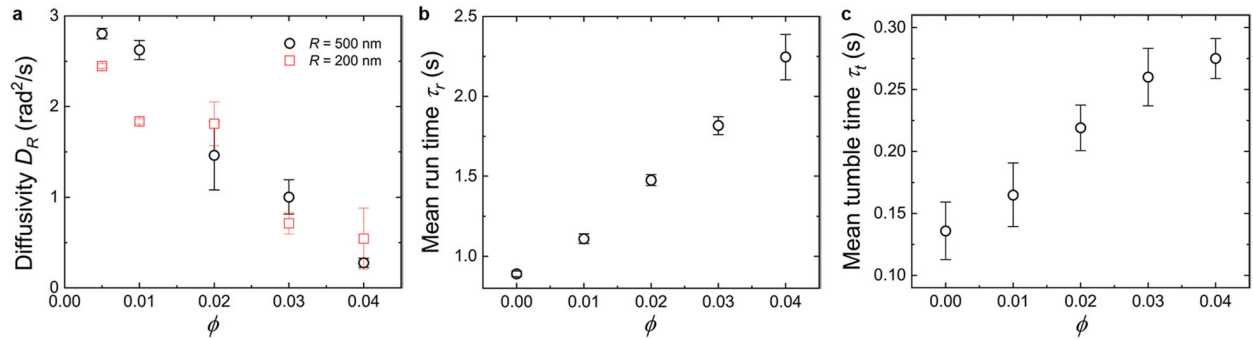

**Extended Data Fig. 2 | The rotational diffusivity (a), the mean run time (b) and mean tumble time (c) of bacteria in dilute colloidal suspensions of increasing concentrations.** Black circles are for suspensions of colloids of radius $R = 500$ nm, whereas red squares are for colloids of $R = 200$ nm. The results are quantitatively similar to those reported in Figs. 3D, 4A, 4B and 4C of Ref. [18].

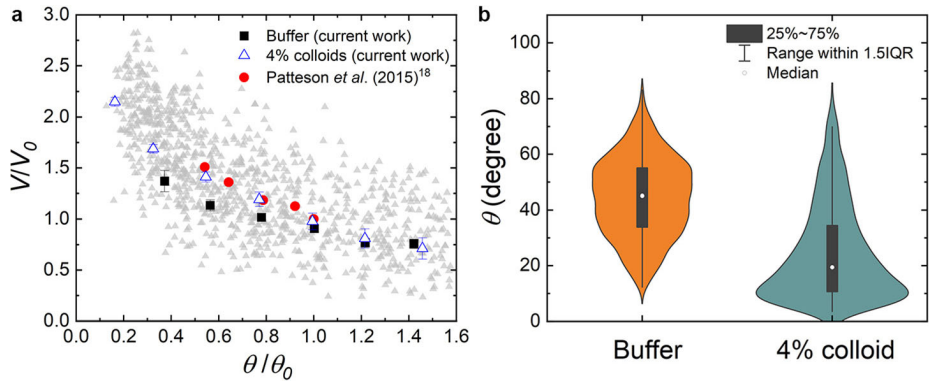

**Extended Data Fig. 3 | Anticorrelation between bacterial wobbling and motility enhancement. a**, The swimming speed versus the wobble angle of individual bacteria in pure buffer (black squares) and in a suspension of colloids of radius $R = 500$ nm and volume fraction $\phi = 4\%$ (blue triangles). The data are obtained by averaging the swimming speed of many bacteria binned over a small range of wobble angles. The raw data for each individual bacteria in the 4% colloidal suspension are also shown as the background (gray triangles). For clarify, we do not show the data of individual bacteria in buffer, which show a similar degree of scattering. $V_0 = 13$ μm/s and $\theta_0 = 45°$ are the average swimming speed and the average wobble angle of bacteria in buffer, respectively. As a comparison, the average swimming speed versus the average wobble of bacteria in polymer solutions of different concentrations from Ref. [18] is also shown (red disks). **b**, A violin plot showing the probability distribution of the wobble angle of bacteria in buffer and in the 4% colloidal suspension. The interquartile range (IQR) gives the difference between the 75th and 25th percentiles of the data. **a** shows that at a given wobble angle, the swimming speed of a bacterium is nearly constant, independent whether it swims in buffer or in colloidal suspensions. **b** shows that the average swimming speed of bacteria increases in colloidal suspensions, because there are more weakly-wobbling bacteria in the suspensions.

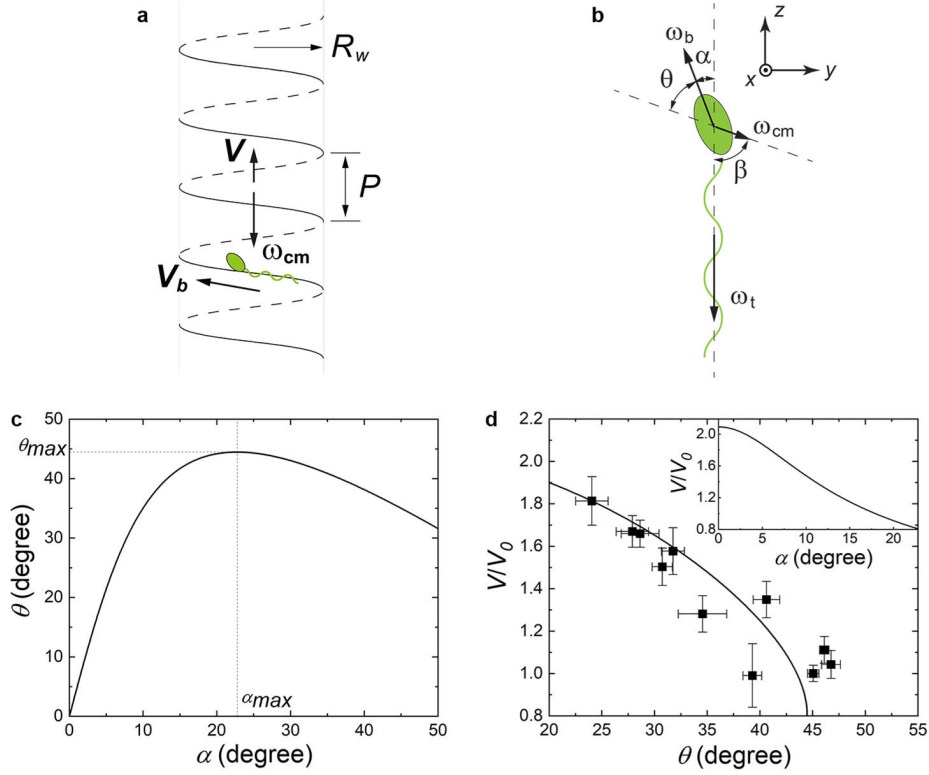

**Extended Data Fig. 4 | Model description and prediction. a**, A schematic showing the 3D helical trajectory of a bacterium. The 2D projection of the 3D trajectory manifests as bacterial wobbling under optical microscopy. The pitch $P$ and the radius $R_w$ of the trajectory are indicated. The velocity of the bacterium tangential to the helical trajectory $V_b$ and the swimming velocity of the bacterium measured in experiments $V$ are also shown. The schematic is not to scale. $R_w$ is comparable to the size of bacteria and much smaller than $P$ for real trajectories (see Extended Data Fig. 5a). The detailed 3D configuration of a bacterium w.r.t. its trajectory is shown in Extended Data Fig. 6. **b**, A schematic showing the motion of a bacterium in our model. The angular velocity of the bacterial body $\omega_b$ and the flagellar bundle $\omega_t$, as well as the angular velocity of the entire bacterium as a rigid body, $\omega_{cm}$, are shown. $\omega_b$, $\omega_t$ and $\omega_{cm}$ are in the same plane, which we define as the $\omega$ plane. The misaligned angle between $\omega_b$ and $\omega_t$, $\alpha$, and the angle of $\omega_{cm}$ with respect to the negative $z$ direction, $\beta$, are also shown. The wobble angle $\theta = \beta - \alpha$. The coordinate system used in our model is defined at the upper right corner. **c**, Wobble angle $\theta$ as a function of the misaligned angle $\alpha$. A maximum wobble angle $\theta_{max} = 44.5°$ reaches when $\alpha_{max} = 22.8°$. **d**, The anticorrelation between the normalized bacterial swimming speed $V/V_0$ and the wobble angle $\theta$. Symbols are from our experiments and the solid line is our model prediction. Figure 3b shows the same results using the normalized wobble angle. In the model, we set $V = 1.8V_0$ at $\theta = 24°$ based on experimental observation. Inset: $V/V_0$ versus the misaligned angle $\alpha$ from our model.

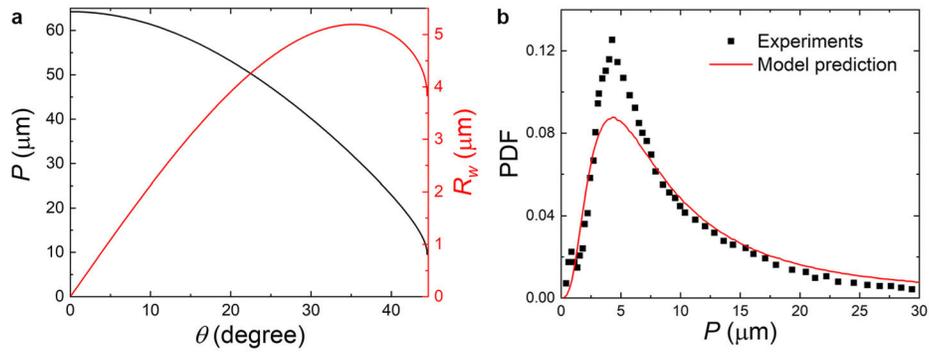

**Extended Data Fig. 5 | Pitch and radius of bacterial helical trajectories. a**, The pitch $P$ and the radius $R_w$ of bacterial helical trajectories as a function of the wobble angle $\theta$. See the definition of $P$ and $R_w$ in Extended Data Fig. 4a. $P$ is shown to the left axis and $R_w$ is shown to the right axis. The limiting pitch of 64 μm can be reached in our model. In comparison, the limiting pitch predicted by the previous minimal model of bacterial wobbling is 4 μm for bacteria with single flagellar bundle[24]. **b**, Probability distribution function (PDF) of the pitch of bacterial helical trajectories. Black squares are experimental data extracted from Fig. 2a of Ref. [24]. The red solid line is our model prediction.

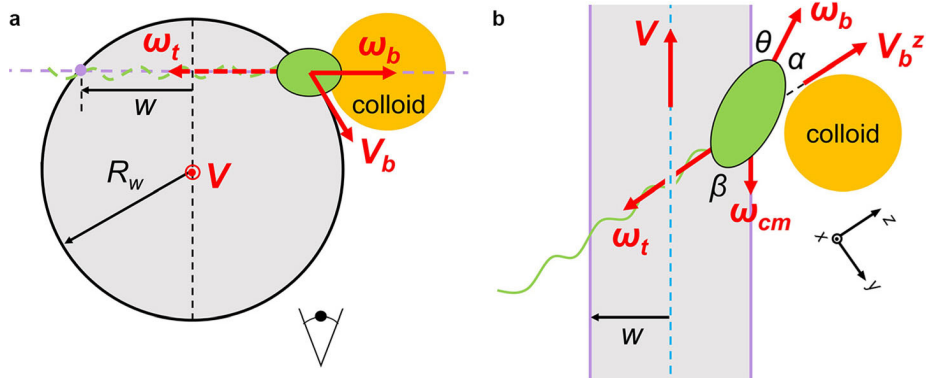

**Extended Data Fig. 6 | Schematics showing the 3D configuration of a swimming bacterium and its helical trajectory. a**, The top view of the configuration. The left-handed helical trajectory encloses a cylindrical space of radius $R_w$ (the gray region). The angular velocity of the body and the flagellar bundle $\omega_b$ and $\omega_t$ are shown. The velocity of the bacterium tangential to the helical trajectory $V_b$ is indicated. Note that $\omega_b$ and $V_b$ tilt above the paper (the solid arrows), whereas $\omega_t$ tilts into the paper (the dashed arrow). The average swimming speed of bacteria measured in experiments $V$ is normal to and points out of the paper. The $\omega$ plane is normal to the paper as indicated by the purple dashed line. The cross-section of the plane with the helical cylinder has a rectangular shape with the width $w < R_w$. The viewpoint of **b** is indicated at the lower right corner. **b**, The side view of the $\omega$ plane. The axis of the helical trajectory is indicated by the vertical dashed line, which is in front of the $\omega$ plane above the paper. The projection of $V_b$ along the direction of $-\omega_{cm}$ gives $V$, whereas the projection of $V_b$ along the direction of the bacterial flagellar bundle $(-\omega_t)$ gives $V_b^z$. The angles $\alpha$, $\beta$ and $\theta$ are the same as those defined in Extended Data Fig. 4b. The coordinate defined in Extended Data Fig. 4b is reproduced on the lower right. As colloids are depleted from the cylindrical space due to the long-range hydrodynamics (Extended Data Fig. 7), colloids that exert torque on the bacterium via the short-range lubrication interaction are outside the cylindrical space. The effect breaks the symmetric role of colloids around the bacterium, which preferably reduces $\alpha$ and therefore suppresses bacterial wobbling.

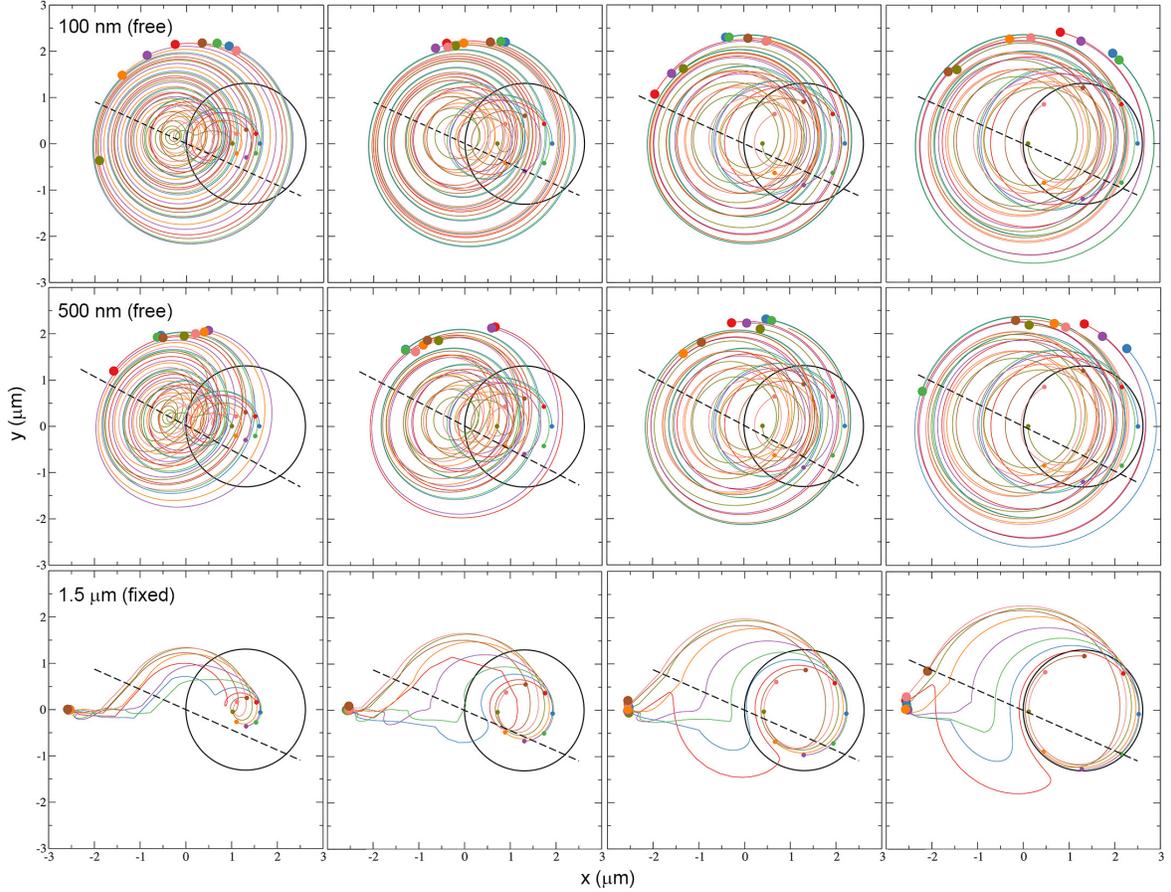

**Extended Data Fig. 7 | Depletion of colloids from the cylindrical space enclosed by the helical trajectory of a wobbling bacteria.** The radius of colloids is 100 nm (top row), 500 nm (middle row) and 1.5 μm (bottom row). While the 100 nm and 500 nm colloids are free, the large 1.5 μm colloids are pinned in the lab frame to match the condition of the optical tweezer experiments. The small disks indicate the initial positions of colloids when the bacterium is far away from the colloids, where the distance between the bacterium and the colloids along the axis of the helical trajectory $-\omega_{cm}$ is $l_{bc} \equiv |(r_b - r_c) \cdot \omega_{cm}|/|\omega_{cm}| = 6.5$ μm. Here, $r_b$ and $r_c$ are the center of bacterial body and the center of the colloids, respectively. The large disks indicate the positions of the corresponding colloids when $l_{bc} = 0$. The lines connecting the small and large disks represent the projection of the 3D trajectories of colloids on the plane normal to $-\omega_{cm}$ in the reference frame of the bacterium, where the bacterial body sits at the origin (0, 0) and orientates in the $\omega$ plane indicated by the black dashed lines. The large black circles mark the boundary of the cylindrical space. The radius $R_w$ and the pitch $P$ of the helical trajectory of the bacterium are 1.3 μm and 6.5 μm, respectively. The radius of bacterial body is 1 μm. All the colloids initially located inside the cylindrical space are depleted out of the space as the bacterium approaches the colloids. The results are obtained from numerical simulations with hydrodynamics interactions[35], where the interaction between one single colloid and one wobbling bacterium is simulated in each simulation run.

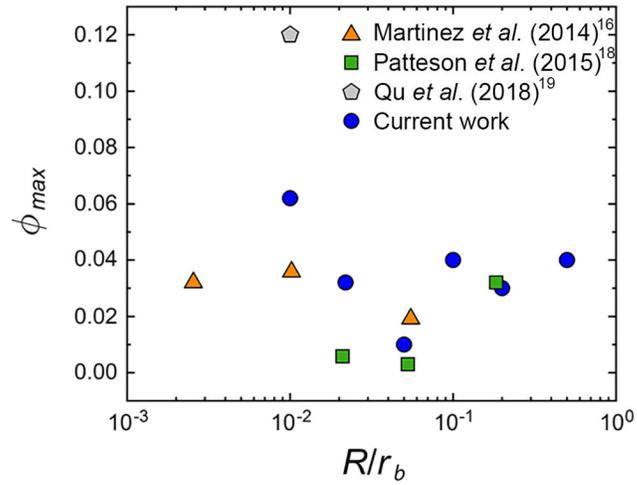

**Extended Data Fig. 8 | The volume fraction at which the motility enhancement peaks, $\phi_{max}$.** $\phi_{max}$ as a function of the radius of colloids or the hydrodynamic radius of polymer coils $R$. $R$ is normalized by the characteristic radius of bacterial body $r_b$ = 1 μm. Except for the datum from Ref. [19], all the $\phi_{max}$ are in the dilute regime.